\documentclass[pre,groupedaddress,showpacs]{revtex4}

\usepackage[dvips]{color}

\usepackage{amsmath}
\usepackage{amssymb}
\usepackage{amsfonts}
\usepackage{mathrsfs}
\usepackage{epsfig}
\usepackage{bm}
\DeclareMathAlphabet{\mathpzc}{OT1}{pzc}{m}{it}

\begin{document}

\title{Dynamics and control of the expansion of finite-size plasmas produced in ultraintense laser-matter interactions}

\author{F. Peano$^{1,3}$, J. L. Martins$^{1}$, R. A. Fonseca$^{1,2}$, and L. O.Silva$^{1}$}
\affiliation{$^{1}$GoLP/Centro de F\'isica dos Plasmas, Instituto Superior T\'ecnico, 1049-001 Lisboa, Portugal}\author{ }
\affiliation{$^{2}$Departamento de Ci\^encias e Tecnologias da Informa\c{c}\~ao, Instituto Superior de Ci\^encias do Trabalho e da Empresa, 1649-026 Lisboa, Portugal}
\author{G. Coppa$^{3}$, F. Peinetti$^{3}$, and R. Mulas$^{3}$}
\affiliation{$^{3}$Dipartimento di Energetica, Politecnico di Torino, 10129 Torino, Italy}

\date{\today}

\begin{abstract}
The strong influence of the electron dynamics provides the possibility of controlling the expansion of laser-produced plasmas by appropriately shaping the laser pulse. A simple irradiation scheme is proposed to tailor the explosion of large deuterium clusters, inducing the formation of shock structures, capable of driving nuclear fusion reactions. Such a scenario has been thoroughly investigated, resorting to two- and three-dimensional particle-in-cell simulations.
Furthermore, the intricate dynamics of ions and electrons during the collisionless expansion of spherical nanoplasmas has been analyzed in detail using a self-consistent ergodic-kinetic model. This study clarifies the transition from hydrodynamic-like to Coulomb-explosion regimes.
\end{abstract}

\maketitle

\newcommand{\ped}[1]{_{\text{#1}}}\newcommand{\api}[1]{^{\text{#1}}}
\newcommand{\diff}[1]{\text{d}#1}
\newcommand{\tc}[2]{\textcolor[rgb]{#1}{#2}}

\section{Introduction}

The rapid developments in laser technology in recent years, both in the IR \cite{CPA} and VUV/X frequency range \cite{Zeitoun}, disclosed new research realms in the field of radiation-matter interactions. Among these, the interaction of ultraintense lasers with jets of atomic or molecular clusters has become a central research topic \cite{Krainov_rep}.
Experiments with laser-cluster interactions have demonstrated tabletop nuclear fusion \cite{Ditmire_Nature,fusion_exp}, opening the way to compact neutron sources; recent studies also suggest the use of the laser-cluster scheme for laboratory investigations of nucleosynthesis reactions, relevant in astrophysical studies \cite{Heidenreich}.

A clustered gas beam can be obtained through adiabatic expansion of a dense gas jet into a vacuum \cite{Hagena,Ter-Avetisyan_PRE}, with the gas backing pressure and the temperature determining the mean cluster size (typically, in the range $1$ nm - $1$ $\mu$m). Such a medium can be regarded as a sparse distribution of tiny solid-like targets, a peculiar configuration that allows for a deep penetration of the laser radiation and a strong laser-matter coupling with many individual, overdense targets: this guarantees an extremely efficient absorption of the radiation (nearly the entire laser energy deposited within a few millimeters propagation length \cite{Ditmire_PRL_3}).

When an ultraintense laser beam hits a cluster, the leading edge of the pulse promptly ionizes the neutral atoms (cf. Ref. \cite{Last_JCP_1} for a detailed analysis of the concurring ionization mechanism in different laser/cluster configurations), forming a dense distribution of ionized matter (usually called a nanoplasma) \cite{Ditmire_PRA_1, Krainov_rep}. Then, the electrons are further heated by the main part of the pulse and the plasma starts to expand, leading to efficient ion acceleration, as first predicted by Dawson [J. M. Dawson, ``On the production of plasmas by giant laser pulses'', Phys. Fluids {\bf 7} pp. 981-987 (1964)]. For small, low-Z clusters (e.g., nm-sized Hydrogen or Deuterium clusters) exposed to extremely intense laser radiation, the electrons can even be completely stripped from the cluster in a few optical cycles, leading to the Coulomb explosion (CE) of the remaining bare-ion distribution. With IR lasers, this happens when the excursion length of the electrons [$c/\omega_{l}\arcsin\left(a_0/(1+a^{2}_{0})\right)$, with $\omega_{l}$ and $a_0$ as central frequency and normalized vector potential of the laser, respectively] is much higher than both the electron skin depth ($c/\omega\ped{pe}$, $\omega_{pe}$ being the electron plasma frequency) and the initial radius of the cluster, $R_0$ \cite{Kishimoto_POP}.
In more general situations, when a relevant fraction of the electrons is bound to the cluster, the expansion process is strongly dependent on the self-consistent dynamics of ions and trapped electrons: when increasing the laser energy or when lowering the cluster size, the expansion conditions vary smoothly from quasi-neutral, hydrodynamic-like regimes to pure CE regimes, as revealed by particle-in-cell (PIC) simulations \cite{Kishimoto_POP,Peano_PhD} of laser-cluster interactions and detailed kinetic analysis of the expansion \cite{Peano_PRL_2}.
This suggests the possibility of exploiting the electron dynamics in order to achieve control over the expansion, which can be done by appropriately shaping the laser pulses. For example, large-scale shock shells (multi-branched structures in the ion phase space, previously predicted by Kaplan, Dubetsky, and Shkolnikov \cite{Kaplan_PRL} for nonuniform CEs) can be driven and controlled by using a proper sequence of two laser pulses having different intensities \cite{Peano_PRL_1}; this opens the way toward intracluster fusion reactions within large Deuterium or Deuterium-Tritium clusters \cite{Kaplan_PRL,Peano_PRA}.

The present paper is organized as follows. After briefly introducing the concept of shock shells in a Coulomb explosion (Sec. \ref{Sec:shocks}), the description of a double-pump irradiation scheme (Sec. \ref{Sec:double-pump}) for inducing and controlling shock shells capable of driving intracluster nuclear reactions (Sec. \ref{Sec:intracluster}) is given, whereas Sec. \ref{Sec:xrays} investigates the possibility of controlling expansions also with VUV/X-ray beams. Section \ref{Sec:Expansion} is devoted to the kinetic analysis of spherical plasma expansions, performed with a novel ergodic-kinetic model (Sec. \ref{Sec:ergodic}): selected results obtained with this model are presented in Sec. \ref{Sec:results}.

\section{Controlled shocks shells}
\label{Sec:part1}

\subsection{Shock shells in Coulomb explosions}
\label{Sec:shocks}

The explosion dynamics of a spherical distribution of cold ions is described by the equation for the radial motion of the ions,
\begin{equation}
	M\frac{\partial^2r\ped{i}}{\partial t^2} = q\ped{i}^2\frac{N\ped{i}(r\ped{i},t)}{r\ped{i}^2} \text{,}
	\label{eq:Newton}
\end{equation}
where $r\ped{i}(r_0,t)$ denotes the trajectory of an ion starting at $t=0$ from radius $r_0$ with zero radial velocity, $M$ is the ion mass, $q\ped{i}=Ze$ is the ion charge, and $N\ped{i}(r,t)$ is the number of ions enveloped at time $t$ by a sphere of radius $r$.
\begin{figure}[!htbp]
\centering \epsfig{file=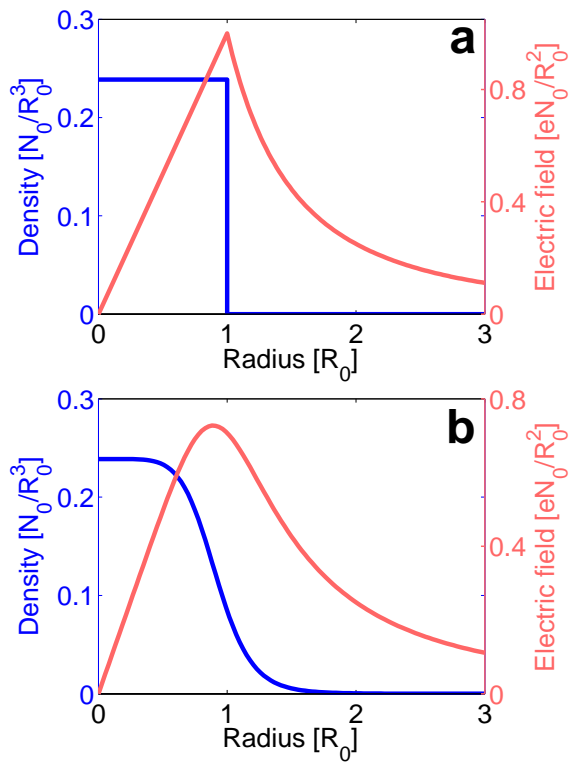, width=3in}
\caption{Initial ion density (blue) and radial electric field (light red) for the case of (a) a uniform, step-like density and of (b) a nonuniform, smoothly-decreasing density, defined as $n\ped{i0}/\left[1+(r_0/R_0)^6\right]^{3/2}$, $R_0$ being the radius of the equivalent uniform sphere.}
\label{fig:uni_vs_nonuni}
\end{figure}
\begin{figure}[!htbp]
\centering \epsfig{file=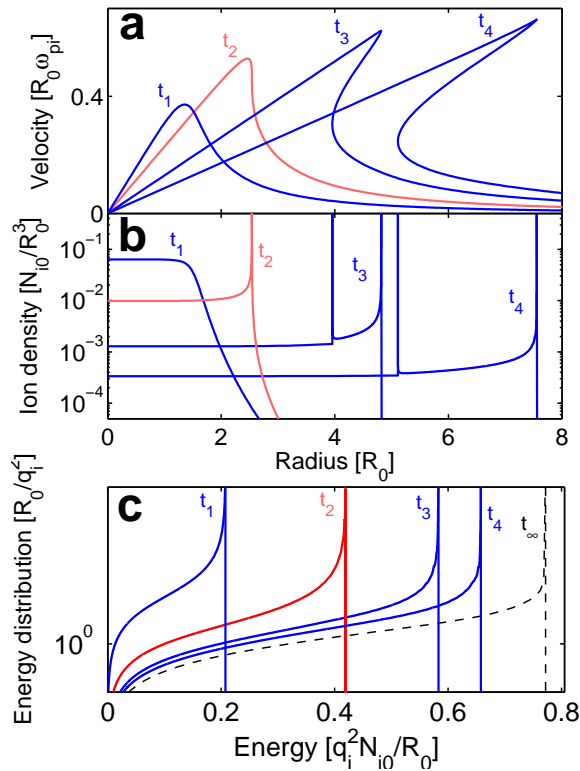, width=3in}
\caption{(Ion (a) phase-space profile, (b) density, and (c) energy spectrum at times $t_1 = 1.97$ $\omega\ped{pi}^{-1}$, $t_2  = 4.23$ $\omega\ped{pi}^{-1}$ (shock formation time), $t_3 = 8.19$ $\omega\ped{pi}^{-1}$, and $t_4 = 12.44$ $\omega\ped{pi}^{-1}$.}
\label{fig:shock}
\end{figure}
When starting from a uniform ion density (equal to $n\ped{i0}$ for $r \leqslant R_0$), the repulsive electric field grows linearly for $r<R_0$, reaching its maximum at the outer boundary (Fig. \ref{fig:uni_vs_nonuni}a), and causing the outer ions to be always faster than the inner ones. In such a particular situation, ions never overtake each other, $N\ped{i}(r,t)$ is conserved along the ion trajectories (so is the total energy of each ion), and Eq. \eqref{eq:Newton} can be integrated analytically \cite{Kaplan_PRL,Parks}: assuming all ions to be initially at rest, the expansion velocity $v\ped{i}(r_0,t)$ is 
\begin{equation}
	v\ped{i} = \sqrt{\frac{2}{3} \ \frac{r\ped{i}-r_0}{r\ped{i}}}r_0\omega\ped{pi} \text{,}
	\label{eq:velocity}
\end{equation}
where $\omega\ped{pi}=\sqrt{4\pi q\ped{i}^2n\ped{i0}/M}$ is the (initial) ion plasma frequency, whereas its radial trajectory is given by
\begin{equation}
	\sqrt{\xi\ped{i} \left(\xi\ped{i}-1\right)} + \log\left(\sqrt{\xi\ped{i}}+\sqrt{\xi\ped{i}-1}\right) = \sqrt{\frac{2}{3}}\omega\ped{pi}t \text{,}
	\label{eq:radius}
\end{equation}
where $\xi\ped{i}=r\ped{i}/r_0$ is the expansion factor. Since the right-hand side of Eq. \eqref{eq:radius} does not depend on $r_0$, $\xi\ped{i}$ is the same for all ions, being a function of time only, i.e., $\xi\ped{i}=\xi(t)$. As a consequence, the $v-r$ phase-space profile is always a straight line [having equation $v=\sqrt{2/3 (\xi-1)/\xi^3}\omega\ped{pi}r$] and the ion density stays always uniform (evolving as $n_0\xi^{-3}$). Correspondingly, the ion energy distribution is $3/2(R_0/q\ped{i}^2)\sqrt{\epsilon/\epsilon\ped{CE}}(1-1/\xi)^{-3/2}$, where $\epsilon\ped{CE}=q\ped{i}^2N\ped{i0}/R_0$ is the asymptotic cutoff energy.
Such solutions could be obtained equivalently from cold-fluid equations ($v$ being now the fluid velocity).
However, when nonuniform density profiles are considered, the kinetic aspects of the process immediately reveal themselves and the features of the expansion change dramatically, as first predicted in Ref. \cite{Kaplan_PRL}, and explored in self-consistent simulations in Refs. \cite{Peano_PRL_1,Peano_PRA}. In fact, if the initial density is a decreasing function of $r$, the repulsive Coulomb field reaches its maximum within the ion sphere \cite{Peano_PhD,Kaplan_PRL} (cf. Fig. \ref {fig:uni_vs_nonuni}b), causing some inner ions to accelerate more than the outer ions in front of them: this ultimately leads to overtaking between ions and the formation of a shock shell. In this situation, the ion trajectories are no longer independent of one another, $N\ped{i}(r,t)$ is not conserved along the ion trajectories (nor is the total energy of each ion), and Eq. \eqref{eq:Newton} no longer admits the analytical solutions reported above. Figure \ref{fig:shock} shows a typical nonuniform expansion (similar to that illustrated in Ref. \cite{Kaplan_PRL}, Fig. 1), starting from the smoothly-decreasing profile of Fig. \ref{fig:uni_vs_nonuni}b:  the phase-space profile starts bending on the right until it becomes multivalued, and a pronounced three-branched shock shell forms (cf. Fig. \ref{fig:shock}a). As predicted by the theory \cite{Kaplan_PRL}, vertical-tangent points in the phase-space profile correspond to a singularity in the ion density (cf. Fig. \ref{fig:shock}b) (such singularity is sometimes called a caustic \cite{Kovalev_JETP}, using a term borrowed from astrophysics \cite{Gurevich}), while horizontal-tangent points correspond to a singularity in the energy distribution (cf. Fig. \ref{fig:shock}c); this appears at the very beginning of the expansion and it is sometimes called the shock predictor \cite{Kaplan_PRL}, since it indicates the presence of a maximum in the velocity profile. As the expansion continues, the shock shell widens radially, involving an ever larger portion of the ion cloud. The physical interest of such pronounced, large-scale shock shells resides in the appearance of large relative velocities within a single exploding cluster, which can lead to energetic ion-ion collisions, thus opening the way to intracluster nuclear fusion reactions \cite{Kaplan_PRL, Peano_PRL_1, Peano_PRA} (e.g., in the case of D or D-T clusters).

\subsection{The double-pump technique}
\label{Sec:double-pump}
\begin{figure}[!htbp]
\centering \epsfig{file=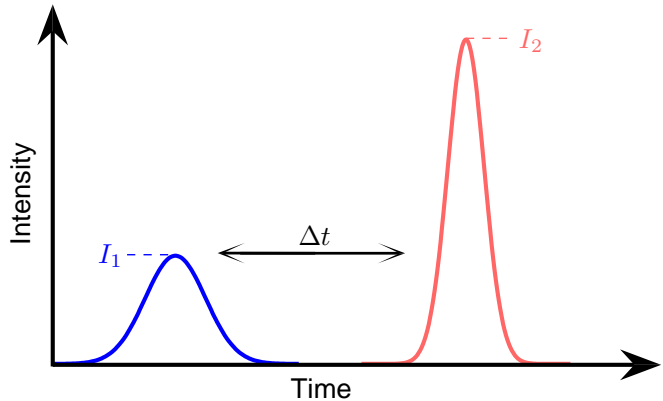, width=3.5in}
\caption{Sketch of the double-pump scheme: the target cluster is first irradiated with a low-intensity laser and, after a delay $\Delta t$, with a high-intensity laser.}
\label{fig:sketch}
\end{figure}

An efficient strategy for producing large-scale shock shells in a controlled fashion is the combination of different expansion regimes. In fact, if a slow, hydrodynamic-like expansion is exploited to provide a decreasing density profile, then a subsequent, abrupt CE naturally leads to the formation of a shock shell, according to the dynamics described in Sec. \ref{Sec:shocks}. Such a slow-expansion/abrupt-explosion dynamics can be obtained rather simply, by irradiating a cluster with two sequential pulses having different intensities, namely, a weaker pulse followed (after a proper time delay) by a much stronger one, as sketched in Fig. \ref{fig:sketch}. The first pulse must be intense enough to ionize the atoms, creating a nanoplasma, but no so intense as to expel a significant fraction of the electrons from the cluster core: in these conditions, a slow expansion takes place, in a quasineutral, hydrodynamic-like regime, and a smoothly decreasing density profile is formed. The second pulse must be extremely intense, in order to strip all the electrons from the cluster, thus driving a CE.
\begin{figure}[!htbp]
\centering \epsfig{file=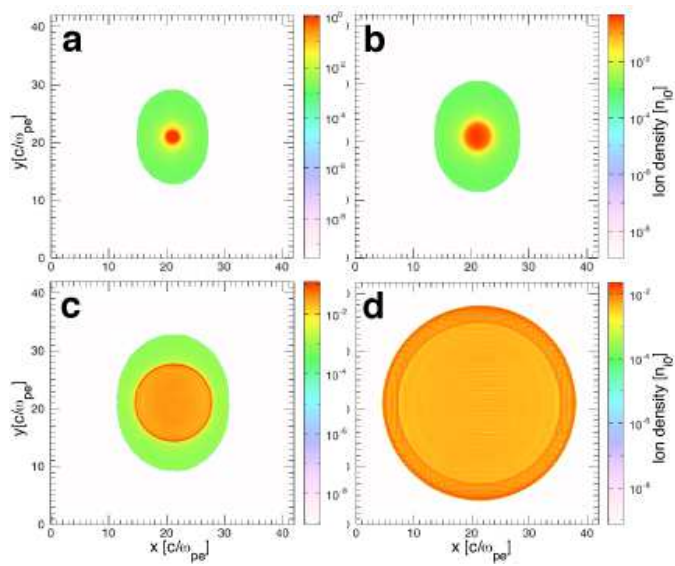, width=3.5in}
\caption{Ion density at (a) $t = 170$ fs, (b) $t = 187$ fs, (c) $t = 206$ fs, and (d) $t = 237$ fs. A $1 \mu\text{m} \times 1 \mu\text{m}$ computational domain has been used; a $840 \times 840$ uniform spatial grid, and $1.2\times 10^6$ particles per species have been employed.}
\label{fig:sim_2D_dens_all}
\end{figure}
\begin{figure}[!htbp]
\centering \epsfig{file=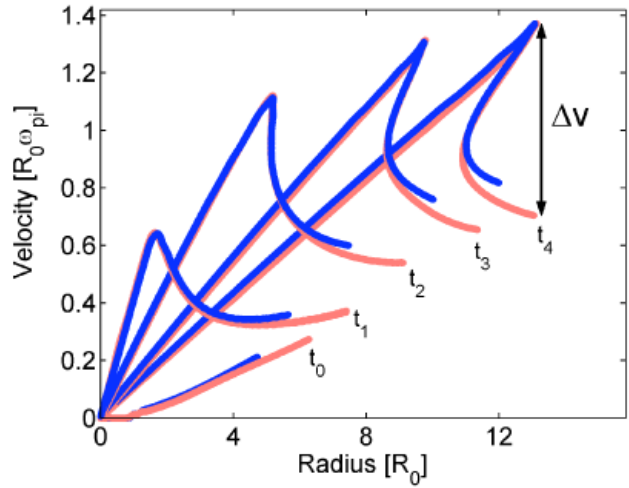, width=3.5in}
\caption{Ion phase-space profile at times $t0 = 170$ fs, $t1 = 187$ fs, $t2 = 206$ fs, $t3 = 225$ fs, and $t4 = 237$ fs, for particles contained in an angle $\Delta \theta = 0.1$ rad around the propagation direction $x$ (blue markers) and the polarization direction $y$ (light-red markers).}
\label{fig:sim_2D_phase}
\end{figure}
The effectiveness of this double-pump technique has been demonstrated \cite{Peano_PRL_1,Peano_PRA} by resorting to 2D and 3D PIC simulations performed using the OSIRIS 2.0 framework \cite{OSIRIS}, closely matching realistic physical scenarios. The results presented in Figs. \ref{fig:sim_2D_dens_all} and \ref{fig:sim_2D_phase} refer to a 2D simulation, in which a deuterium cluster (initial radius $R_0=32$ nm, density $n_0 = 4.56\times 10^{22}$ $\text{cm}^{-3}$) is first irradiated by a relatively weak laser pulse (peak intensity $1\times 10^{16}$ $\rm{W}/\rm{cm}^2$, duration 35 fs, wavelength $820$ nm) and, after a $170$ fs delay, it is irradiated by a second, extremely intense pulse (peak intensity $2.5\times 10^{19}$ $\rm{W}/\rm{cm}^2$, duration $20$ fs, wavelength $820$ nm); both pulses propagate in the $x$ direction, being linearly polarized along the $y$ direction. Figures \ref{fig:sim_2D_dens_all} and \ref{fig:sim_2D_phase} show the evolution of the ion density distribution and the ion phase-space profile, respectively, during the CE stage. When the second pulse hits the plasma sphere (cf. Fig. \ref{fig:sim_2D_dens_all}a), the outer ions are expanding slowly, under the pressure of the hot electrons, while the cluster core is still dense, with the inner ions still at rest (cf. Fig. \ref{fig:sim_2D_phase}). As a result, when the electrons are suddenly swept away by the second laser, the inner ions experience a much higher repulsive force than the outer ions do, so that they rapidly overrun them, leading to the formation of a large, pronounced shock shell, clearly visible in both ion density (Fig. \ref{fig:sim_2D_dens_all}a) and phase space (Fig. \ref{fig:sim_2D_phase}). In this example, the relative velocities between ions belonging to different branches in the shock shell would allow for collision energies as high as $200$ keV, which is enough for efficient DD fusion to occur.

\subsection{Fusion reactions within single exploding clusters}
\label{Sec:intracluster}
The ability of generating pronounced shock shells involving high relative velocities within single exploding clusters makes the phenomenon attractive as a possible way to induce nuclear reactions. As already pointed out in Ref. \cite{Kaplan_PRL}, the rates for intracluster reactions can be much higher than those for intercluster reactions, because the typical densities within a shock shell ($\lesssim n\ped{i0} \sim 10^{22}$ cm$^{-3}$) can be much higher than those within the hot plasma filament resulting from the exploded clusters, whose density is  $\sim 10^{19}$ cm$^{-3}$. However, it is crucial to consider also that a shock shell stays appreciably dense only for a very brief time ($\sim 10$ fs), much shorter than the typical disassembly time of the plasma filament ($\sim 10-100$ ps). In fact, as shown in Ref. \cite{Peano_PRA}, the intracluster-reaction rates exhibit a sharp, time-resolved peak right after the shock-shell formation. 
In order to compute the actual number of shock-driven, intracluster fusion reactions, it is necessary to sum up all possible contributions from collisions between ions belonging to different velocity branches. The number of reactions per unit time and volume, $\mathcal{R}$, is given by
\begin{equation}
\mathcal{R} = \sum_{h<k}
                         n_{h}(r)
                         n_{k}(r)
                         \sigma \left( | v_{h} - v_{k}|  \right)
                         |v_{h} - v_{k}|
                         \text{,}
\label{eq:rrate1}
\end{equation}
where $\sigma$ is the cross section for DD fusion, while $n_{h}$ and $v_{h}$ indicate, respectively, the ion density and velocity on the $h$th branch. The number of intracluster fusion reactions, $\mathcal{N}$, is obtained by integrating $\mathcal{R}$ over time and space, as $\mathcal{N}= 4\pi \int_{t\ped{sh}}^{+\infty}  \int_{r\ped{sh}}^{R\ped{sh}} \mathcal{R} {r}^{2}\diff{r} \diff{t}$, where $r\ped{sh}$ and $R\ped{sh}$ are the shock shell boundaries, whereas $t\ped{sh}$ is the shock formation time. Clearly, $\mathcal{N}$ is strongly dependent on the dynamics of the shock shell, which can be tailored by varying the delay between the two laser pulses as well as their intensities. A detailed analysis of the influence of the double-pump parameters on the intracluster fusion rate (cf. Ref. \cite{Peano_PRA}) reveals that optimal delay/intensity combinations exist that maximize $\mathcal{N}$. For very large clusters ($R_0 \gtrsim 100$ nm) and optimal double-pump configuration, the intracluster reaction yield can become comparable with the interluster neutron yield, with $\sim 10\%$ of the fusion reactions arising from intracluster collisions \cite{Peano_PRA}. 
Since intercluster reactions occur on a much longer time scale than intracluster ones ($\mathcal{R}$ is sharply peaked in time \cite{Peano_PRA}), a well tuned double-pump experiment should provide a clear signature for the occurrence of intracluster, shock-driven fusion reactions, in the form of a time-resolved burst of fusion neutrons anticipating the usual bulk of fusion neutrons due to intercluster reactions.

\subsection{Expansion control with intense VUV/X-ray beams}
\label{Sec:xrays}
The possible use of the double-pump technique also with VUV/X-ray radiation is currently being investigated. With very short wavelengths, the expulsion of electrons from the cluster, which is necessary for a Coulomb explosion to occur, depends essentially on the ionization dynamics and much less on the electron motion in the radiation field [since now the electron excursion length ($\sim c/\omega_{l}$) is very small compared to the typical cluster radius]. Thus, the full details of the ionization processes involved in this situation (e.g., photoionization, Auger decay, collisional ionization, and strong field ionization) must be considered. For example, when the main ionization process is the photoelectric effect (as in D clusters), the mean energy acquired by the electrons is approximately equal to the photon energy, $\epsilon_{\gamma}=h\nu$. A Coulomb explosion can occur if the number of photons hitting the cluster is sufficient to ionize most of the atoms, and $\epsilon_{\gamma}$ is higher than the electrostatic potential well generated by the ions. This suggests that a double-pump technique can still be employed to drive a shock shell, even though the physical mechanisms it relies on are different. In fact, the goal of the first, weak pulse is now to ionize a small fraction of the atoms in the cluster, in order to drive a slow expansion, whereas the second, strong pulse must ionize all the  atoms remaining in the cluster core, causing a sudden explosion. In these conditions, a shock shell forms because some of the ions created by the second pulse overtake the slowly expanding ions, provided that appropriate radiation intensities and delay between pulses are employed.
\begin{figure}[!htbp]
\centering \epsfig{file=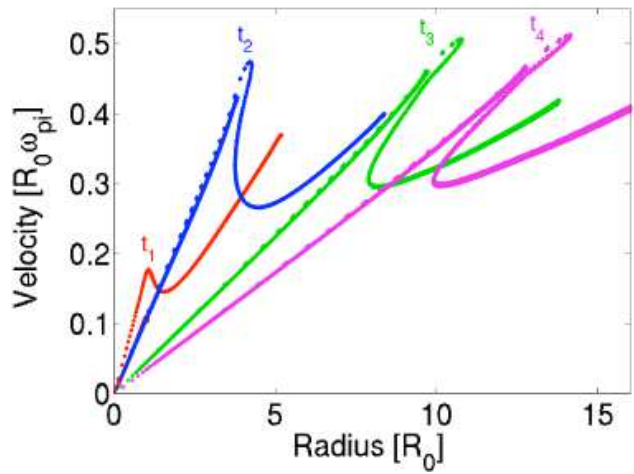, width=3.5in}
\caption{Ion phase-space profile at times $t_1 = 160$ fs, $t_2 = 250$ fs, $t_3 = 390$ fs, and $t_4 = 460$ fs. The computational domain is a cube with a 1.2 $\mu$m side (256x256x256 cells), and 125 particles per cell per species have been used (in the region initially occupied by the cluster).}
\label{fig:xdouble}
\end{figure}
In order to test the effectiveness of this technique, PIC simulations have been used \cite{OSIRIS}, which describe the evolution of the cluster given the initial distributions for the electron and ion populations. This approach is valid whenever the ionization time is short with respect to the expansion time scale, which is usually a good approximation for the intense light sources available in the VUV/X-ray energy range. The formation of a pronounced shock shell, obtained using this technique, is shown in Fig. \ref{fig:xdouble}, referring to an ideal situation in which it is assumed that a deuterium cluster (initial radius $R_0=30$ nm, density $n_0 = 10^{22}$ cm$^{-3}$) undergoes ionization processes which lead to the formation of a population of $N_0/3$ electrons (where $N_0$ is the number of initially neutral atoms) with 12.4 keV energy, followed by a second population of $2N_0/3$ electrons with equal energy, after a 150 fs delay. We observe that at ultra high intensities in XUV/X-ray sources the ionization dynamics is poorly understood since strong field ionization and stabilization against this mechanism may come into play \cite{Reiss}. The balance between the photoelectric, strong field ionization and stabilization processes must be analyzed in order to obtain the correct picture for the energy distribution of the electrons (which actually determines the dynamics of the explosion). The analysis of double-pump scenarios, taking into account the full dynamics of ionization, with beam configurations available at LCLS and XFEL and with different cluster constituents will be presented in future publications.

\section{Kinetic analysis of spherical plasma expansions}
\label{Sec:Expansion}

The possibility of controlling the expansion by acting on the amount of energy transferred from the radiation to the electrons stresses the need for a deep understanding of the physics of the electron-driven collisionless expansion of spherical, nm-sized plasmas, composed of cold ions and hot electrons. This is also crucial for particular applications, such as the biomolecular imaging with ultrashort X-ray pulses \cite{xrays}, where sample damage before the imaging time must be avoided.

Since the first experiments on laser-cluster interactions were performed \cite{fusion_exp}, it has been debated \cite{Ditmire_PRA_1,hydrodynamic} whether the cluster expansions were driven mainly by the pressure of the hot electrons (hydrodynamic-like regime) or by the strong repulsive field which forms when a large fraction of the electrons leave the ion core (CE). Actually, these two expansion regimes are just two opposite limits resulting from the same physical process, namely the formation of a repulsive electric field due to energetic electrons trying to escape from the cold-ion core. When the kinetic energy of the electrons is very small compared to the electrostatic energy stored in the ion core, the electric field is localized at the ion front \cite{Ditmire_PRA_1,Fukuda,Mora}, the whole cluster being approximately neutral, and the expansion is hydrodynamic-like; on the contrary, when the kinetic energy of the electrons is much higher than the ion electrostatic energy, most of the electrons are free to escape and the CE of the ion sphere takes place. However, in a general scenario, the dynamics will be a mixture of the phenomenology of the two limits, and a kinetic description is necessary in order to capture the detailed features of each particular regime.

\subsection{Ergodic model}
\label{Sec:ergodic}
In order to analyze accurately the dynamics of the expansion in any intermediate situation, we have developed a new kinetic model, which allows for a clear identification of the transition from the hydrodynamic expansion to the CE regime, thus setting the range of validity for the CE approximation \cite{Peano_PRL_2}.
Based on the assumptions that the electrons are nonrelativistic and that the time scale for their motion is much shorter than the time scale for the ion motion, the model describes the expansion dynamics self-consistently by following the radial motion of the ions, whereas treating the evolution of the electron distribution as a sequence of equilibrium configurations, represented by stationary solutions of the Vlasov equation for the electrons. In the case of ideal, spherical symmetry, the general stationary solution is a function of the Hamiltonian $\mathscr{H}$ and of the absolute value of the angular momentum, $\ell$, which are constants of motion of the system. However, and by noticing that possible perturbations to the spherical symmetry would rapidly break the $\ell$-conservation, while leaving the $\mathscr{H}$-conservation unaltered, the distribution function in equilibrium can be assumed to depend on $\mathscr{H}$ only, i.e., the system can be assumed to be ergodic. Accordingly, the electron density in equilibrium with the electrostatic potential $\Phi$ can be written as
\begin{equation}
n\ped{e}(r) =\dfrac{1}{4\pi r^2}\int\rho\left(\epsilon\right)
\mathscr{P}\left(r,\epsilon;\{\Phi\}\right)\diff\epsilon \text{,}
\label{eq:ne}
\end{equation}
where $\rho$ is the energy distribution of the electrons and
\begin{equation}
\mathscr{P}\left(r,\epsilon;\{\Phi\}\right) = \dfrac{ r^2\left( \epsilon+e\Phi\right)^{\frac{1}{2}} }
{ \displaystyle\int {r^{\prime}}^2
\left[ \epsilon+e\Phi\left(r^{\prime}\right)\right]^{\frac{1}{2}}\diff r^{\prime}}
\label{eq:P}
\end{equation}
is the probability of finding an electron having total energy $\epsilon$ at the radial position $r$. The self-consistent potential $\Phi$ satisfies the Poisson equation $\nabla^2\Phi = 4\pi e(n\ped{e}-n\ped{i})$, which is nonlinear because $n\ped{e}$ depends on $\Phi$. In order to obtain a closed set of equations describing the plasma expansion, one still needs a rule to update the equilibrium configuration when the ion motion is considered. Since the variations of $\Phi$ induced by the ion motion are slow when compared with the time scale for the electron motion, the theory of adiabatic invariants for time varying Hamiltonians is used. Here, the appropriate invariant is the ergodic invariant (cf., for example, Ref. \cite{Ott})
\begin{equation}
\mathcal{I}\left(\epsilon\right) = \frac{32\sqrt{2}}{3}\pi^2m\ped{e}^{3/2} 
\displaystyle\int \left(\epsilon+e\Phi\right)^{\frac{3}{2}} r^2 \diff r \text{,}
\label{eq:I}
\end{equation}
which is the phase-space volume enclosed by the surface $\frac{1}{2}m\mathbf{v}^2-e\Phi=\epsilon$. The conservation of $\mathcal{I}$ means that the energy variation can be expressed as the ensemble average of the potential-energy variation, thus providing the desired rule for the evolution of the electron energy.
Finally, a closed set of equations can be written in the form
\begin{align}
	&M \dfrac{\partial^2 r\ped{i}}{\partial t^2} = -Ze \dfrac{\partial \Phi}{\partial r}(r\ped{i})
	\tag{8a}\label{eq:model_a}\\
	&\dfrac{1}{r^2}\dfrac{\partial}{\partial r}\left(r^2\dfrac{\partial \Phi}{\partial r}\right) = 4\pi 				
	e\left(n\ped{e}-Zn\ped{i}\right)
	\tag{8b}\label{eq:model_b}\\
	&n\ped{i}(r\ped{i}) = n\ped{i0}(r_0)\dfrac{r_0^2}{r\ped{i}^2}\!\Big/\dfrac{\partial r\ped{i}}{\partial r_0}
	\tag{8c}\label{eq:model_c}\\
	&n\ped{e} = \dfrac{1}{4\pi r^2}\displaystyle\int \rho_0(\epsilon_0)
	\mathscr{P}\left(r,\epsilon;\{\Phi\}\right)\diff{\epsilon_0}
	\tag{8d}\label{eq:model_d}\\
	&\dfrac{\diff\epsilon}{\diff t} = -e\displaystyle\int \dfrac{\partial\Phi}{\partial t} 	
	\mathscr{P}\left(r,\epsilon;\Phi\right)\diff r
	\tag{8e}\label{eq:model_e}
\end{align}
where Eqs. \eqref{eq:model_a} and \eqref{eq:model_e}, determining the evolution of the ion trajectories $r\ped{i}(r_0,t)$ and the electron energies $\epsilon(\epsilon_0,t)$ ($\epsilon_0$ being the initial energy), are coupled with the nonlinear Poisson equation, which provides the self-consistent $\Phi(r,t)$ (here, $\Phi$ is set to zero at infinity, so that $\epsilon<0$ for trapped electrons). In Eq. \eqref{eq:model_d}, the relation $\rho(\epsilon,t)\diff{\epsilon}=\rho_0(\epsilon_0)\diff{\epsilon_0}$ has been used, $\rho_0(\epsilon_0)$ being the initial energy distribution of the electrons.
Equation \eqref{eq:model_c} expresses the ion density in terms of the Lagrangian coordinate $r_0$, under the hypothesis of no ion overtaking ($\partial r\ped{i}/\partial r_0 \neq 0$) \cite{Kaplan_PRL} (when shock shells are present, a  more complicated expression has to be used \cite{Peano_PhD}).
The solution of Eqs. (8) requires the knowledge of $n\ped{i0}(r_0)$ and $\rho_0(\epsilon_0)$.
Depending on the physical problem considered, the determination of $\rho_0(\epsilon_0)$ can be nontrivial. This is precisely the case when considering expansions starting from an initially neutral distribution ($n\ped{e} = Zn\ped{i}$, and $\Phi=0$ everywhere) with hot Maxwellian electrons, having temperature $T_0$, because such configuration is far from equilibrium.
In that case, the plasma expansion can be regarded as a sequence of two distinct processes: first, a rapid expansion of the electrons is observed, which leads to a VP equilibrium configuration before the ions move appreciably; second, a slow expansion of the plasma bulk is observed. While the latter stage can be certainly analyzed using Eqs. (8), a different procedure must be used to determine the equilibrium configuration following the first stage [in order to obtain the correct form of $\rho_0(\epsilon_0)$]. To this purpose, the actual charging transient described by the full VP model is replaced by a virtual charging transient, in which an external potential barrier, initially confining the electrons, is gradually moved from $R_0$ to infinity, with a series of small radial displacements. Each time the barrier is moved farther (from $R\ped{w}$ to $R\ped{w}+\delta R\ped{w}$), the new self-consistent potential ($\Phi+\delta\Phi$) and electron energy ($\epsilon+\delta\epsilon$) have to be calculated. In order to mimic an expansion into a vacuum (which the real transient is), direct energy exchanges between the electrons and the expanding barrier (expansion work) must be avoided. This is accomplished by expressing the energy variation $\delta \epsilon$ as
\begin{equation}
	\delta\epsilon = -e\int^{R\ped{w}}_{0} \delta\Phi \mathscr{P}\left(r,\epsilon;\Phi\right)\diff{r}
	\text{.}
\tag{9} \label{eq:delta_e}
\end{equation}
Although formally identical to Eq. \eqref{eq:model_e}, Eq. \eqref{eq:delta_e} is not equivalent to the conservation of the ergodic invariant (which would  bring on an extra term in the expression for $\delta \epsilon$, accounting for the expansion work). The variations $\delta \Phi$ and $\delta\epsilon$ are determined by simultaneously solving Eqs. \eqref{eq:model_b} and \eqref{eq:delta_e} with a suitable iterative scheme. 
Once coupled with this procedure, which provides the initial conditions $\rho_0(\epsilon_0)$, Eqs. (8) constitute a self-consistent model describing the collisionless expansion of spherical nanoplasmas, starting from an initially neutral distribution composed of cold ions and hot, Maxwellian electrons. This model has proved to be suitable for a detailed analysis of the expansion mechanism in a wide range of conditions, since the dynamics can be determined by the single dimensionless parameter $\hat{T}_0 = Zk\ped{B}T_0/\epsilon\ped{CE}=3\lambda\ped{D}^2/R_0^2$ ($\lambda\ped{D}$ being the initial Debye length for the electrons), which accounts for both the initial electron temperature and the cluster parameters (size and density).

\subsection{Selected results}
\label{Sec:results}
In the following, a detailed analysis of the expansion dynamics is presented for two reference cases, (a) $\hat{T}_0 = 7.2 \times 10^{-3}$ and (b) $\hat{T}_0 =7.2 \times 10^{-2}$ (for the cluster size and the density employed in Sec. \ref{Sec:double-pump}, these values correspond to $T_0 = 2$ keV and $20$ keV, respectively), also illustrating the whole-range $\hat{T}_0$-dependence of the key expansion parameters.

\begin{figure}[!htbp]
\centering \epsfig{file=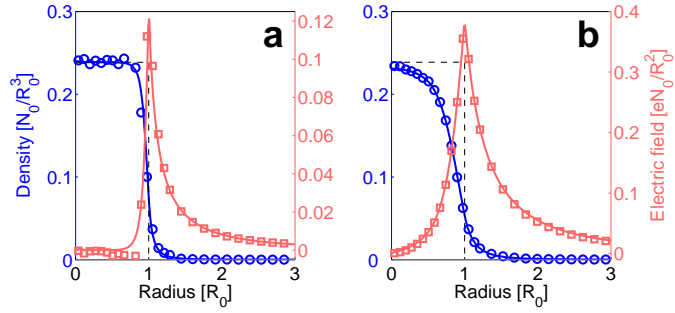, width=3.5in}
\caption{Equilibrium electron density (blue) and electric field (light red), after the charging transient with immobile ions, for (a) $\hat{T}_0= 7.2\times 10^{-3}$ and (b) $\hat{T}_0= 7.2\times 10^{-2}$. Lines refer to the ergodic model and markers to PIC simulations.}
\label{fig:n_E_all}
\end{figure}
\begin{figure}[!htbp]
\centering \epsfig{file=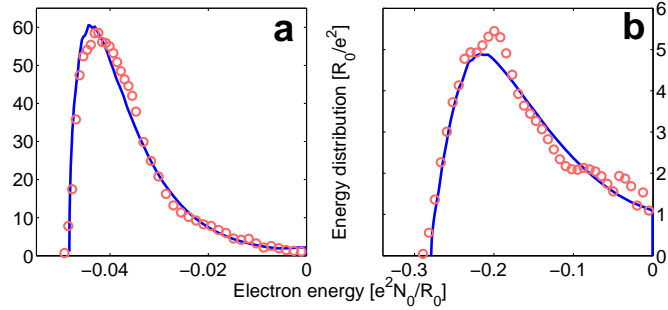, width=3.5in}
\caption{Equilibrium electron energy distribution $\rho_0(\epsilon_0)$, after the charging transient with immobile ions, for (a) $\hat{T}_0= 7.2\times 10^{-3}$ and (b) $\hat{T}_0= 7.2\times 10^{-2}$. Lines refer to the ergodic model and markers to OSIRIS simulations.}
\label{fig:rho_E_all}
\end{figure}
Figures \ref{fig:n_E_all} and \ref{fig:rho_E_all} show the self-consistent equilibrium configuration of the electrons after the initial charging transient (comparisons with reference results obtained with OSIRIS are also shown, which confirm the validity of the barrier method, outlined before, to obtain the conditions after the charging transient). In Fig. \ref{fig:n_E_all}, the electron density is plotted, along with the corresponding electric field: the positive charge buildup at the ion front, $\Delta Q$, is in (a) $12.5\%$ and in (b) $38\%$ of the total ionic charge $eN_0$. Figure \ref{fig:rho_E_all} shows the equilibrium energy distribution $\rho_0(\epsilon_0)$, to be used as initial condition for the bulk expansion.
\begin{figure}[!htbp]
\centering \epsfig{file=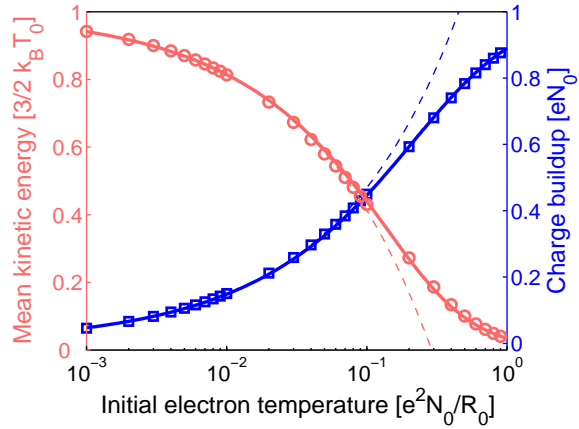, width=3in}
\caption{Charge buildup $\Delta Q$ (blue) and mean kinetic energy of trapped electrons, $\mathcal{E}$, (light red) as functions of $\hat{T}_0$, after the charging transient with immobile ions. Circles refer to the ergodic model, solid lines to the fit laws of Eqs. \eqref{eq:fit1} and \eqref{eq:fit2}. Dashed lines show the power-law behavior of $\Delta Q$ and $\mathcal{E}$ for $\hat{T}_0 \ll 1$.}
\label{fig:Qeq_Teq}
\end{figure}
The $\hat{T}_0$-dependence of $\Delta Q$ and of the mean kinetic energy of the trapped electrons, $\mathcal{E}$, is displayed in Fig. \ref{fig:Qeq_Teq}. Simple, accurate fits exist for these quantities, as
\begin{equation}
	\frac{\Delta Q}{eN_0} = \mathcal{F}_{2.60}\left(\sqrt{6/e}\hat{T}_0^{1/2}\right)
	\text{,}
\tag{10} \label{eq:fit1}
\end{equation}
\begin{equation}
	\frac{\mathcal{E}}{\frac{3}{2}k\ped{B}T_0}=
	1-\mathcal{F}_{3.35}\left(1.86\hat{T}_0^{1/2}\right)
	\text{,}
\tag{11} \label{eq:fit2}
\end{equation}
where $\mathcal{F}_{\mu}(x) = x/(1+x^{\mu})^{1/\mu}$. For $\hat{T}_0 \ll 1$ (i.e. $\lambda_0 \ll R_0$), Eq. \eqref{eq:fit1} reduces to $\Delta Q/Q_0 \simeq \sqrt{6/e} \hat{T_0}^{1/2}$, thus recovering the theoretical results for planar expansions \cite{Mora,Crow}, while Eq. \eqref{eq:fit2} reduces to $\mathcal{E}/(\frac{3}{2} \ k\ped{B}T_0) \simeq 1 - 1.86\:\hat{T}_0^{1/2}$.
\begin{figure}[!htbp]
\centering \epsfig{file=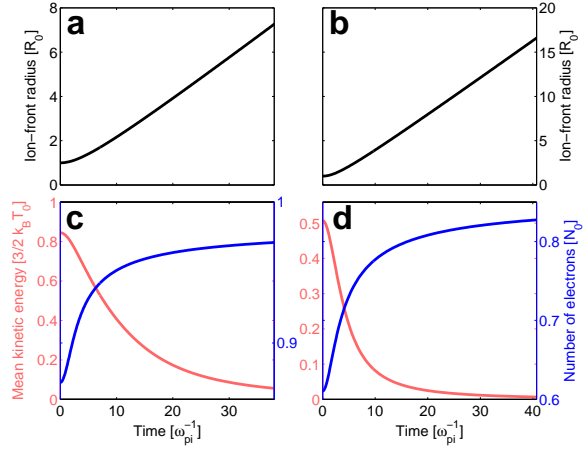, width=3in}
\caption{Evolution of the ion-front trajectory (panels a and b), of the electronic charge enveloped by the ion front, $eN_0 - \Delta Q$, (panels c and d, blue) and of the mean kinetic energy of trapped electrons, $\mathcal{E}$, (panels c and d, light red), for (a,c) $\hat{T}_0= 7.2\times 10^{-3}$ and (b,d) $\hat{T}_0= 7.2\times 10^{-2}$.}
\label{fig:QTR_all}
\end{figure}
The time evolution of $\Delta Q$ and $\mathcal{E}$ during the bulk expansion is illustrated in Fig. \ref{fig:QTR_all}: as the ions expand, gaining kinetic energy, the electrons cool down and the charge buildup decreases (until a ballistic regime is reached for both species \cite{Manfredi}). Such behavior of the electrons strongly affects the ion dynamics and their resulting energy spectrum, since the asymptotic energy $\epsilon_\infty$ of an ion is given by
\begin{equation}
\frac{\epsilon_\infty(r_0)}{Ze} = \frac{q(r_0,0)}{r_0} + \int^{\infty}_{0} \! \!  \frac{1}{r\ped{i}(r_0,t)}\frac{\partial q\left(r\ped{i}(r_0,t),t\right)}{\partial t}\diff{t}
\tag{12} \label{eq:E_ion} \text{,}
\end{equation}
\begin{figure}[!htbp]
\centering \epsfig{file=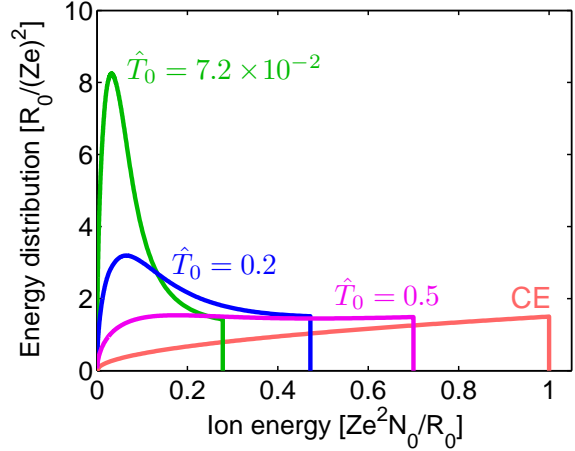, width=3in}
\caption{Asymptotic ion energy spectra for different values of $\hat{T}_{0}$, compared with the theoretical asymptotic spectrum for the CE case.}
\label{fig:ionspectrum_T0}
\end{figure}
\begin{figure}[!htbp]
\centering \epsfig{file=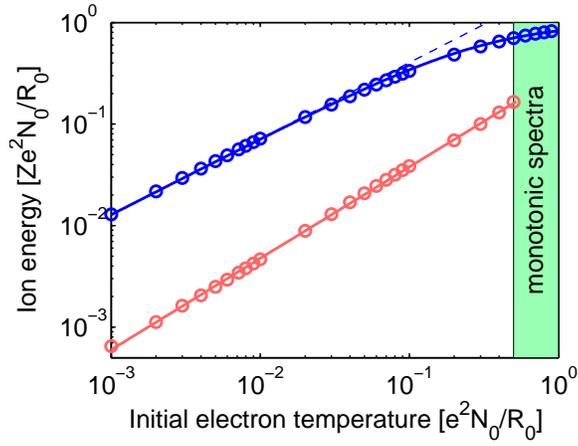, width=3in}
\caption{Cutoff ion energy (blue) and location of the maximum in the ion energy spectrum (light red) as functions of $\hat{T}_0$: circles refer to the ergodic model, solid lines to the fit laws in the text. The dashed line represents the power-law behavior of $\epsilon\ped{max}$ for $\hat{T}_0 \ll 1$.}
\label{fig:E_max}
\end{figure}
where $q(r,t)$ is the net charge buildup enveloped by a sphere of radius $r$ at time $t$. The integral term (vanishing for a CE) accounts for the energy loss due to the decrease of the positive charge buildup experienced by each ion along its trajectory.
Figure \ref{fig:ionspectrum_T0} shows the asymptotic ion spectrum for different values of $\hat{T}_0$. For $\hat{T}_0 < 0.5$, profound differences exist with respect to the CE case, with the distribution exhibiting a local maximum far from the cutoff energy. The spectrum is monotonic only for $\hat{T}_0 > 0.5$ (a condition already close to the CE regime, with cutoff energy above $0.7\epsilon\ped{CE}$). Therefore, the transition value $\hat{T}_0 = 0.5$ can be considered as the lower bound for the validity of the CE approximation.
The cutoff energy $\epsilon\ped{max}$ and the location of the maximum, $\epsilon\ped{peak}$, are plotted in Fig. \ref{fig:E_max} as functions of $\hat{T}_0$. The cutoff energy can be accurately fitted as
\begin{equation}
\epsilon\ped{max}=\mathcal{F}_{1.43}\left(2.28\:\hat{T}_0^{3/4}\right)\epsilon\ped{CE}\text{,}
\tag{13} \label{eq:fit3}
\end{equation}
which, for $\hat{T_0}\ll 1$, reduces to $\epsilon\ped{max}\simeq 2.28\:\hat{T}_0^{3/4}\epsilon\ped{CE}$, while, for $\hat{T}_0 < 0.5$, $\epsilon\ped{peak}$ exhibits the simple power-law behavior $\epsilon\ped{peak}=0.3\hat{T}_0^{0.9}\epsilon\ped{CE}$. These formulae provide accurate scaling laws, valid for any combination of $R_0$, $n_0$, and $T_0$ (as long as relativistic effects can be neglected), which can be useful when interpreting experimental data \cite{Peano_PRL_2,Sakabe}. 

\section{Conclusions}
The study presented here demonstrates the possibility of controlling the expansion of finite-size plasmas with appropriately shaped laser pulses. In fact, by irradiating a spherical target with sequential beams (double-pump technique) the phase-space dynamics of the ions can be manipulated, driving large-scale shock shells. These peculiar kinetic structures can lead to highly energetic ion-ion collisions and intracluster nuclear fusion reactions (in the case of clusters composed of deuterium or mixtures of deuterium and tritium).
The features of the shock shells can be tailored by varying the intensity of the two laser pulses and the time delay between them, thus providing an efficient way to control the explosion of nanometer-sized plasmas. In the future, the availability of even more intense lasers will allow the control of the expansion of micrometer-sized, solid spheres, thus opening the way toward ultraprompt, ultralocalized sources of nuclear reactions, which can be useful for neutron production or nucleosynthesis.

Since expansion control is achieved by acting on the electron dynamics, a detailed knowledge of the self-consistent dynamics of ions and electrons is required. The ergodic model described here provides a new, efficient tool, capable of describing the kinetics of spherical nanoplasma expansions with high accuracy.
Although the ergodic model has been employed here in the paradigmatic case of initially Maxwellian electrons, it can also be used with arbitrary initial distributions of electron energy; this can be of interest in particular physical scenarios where accurate control over the expansion is crucial, such as in single-shot X-ray imaging of biological samples, in order to guarantee that no relevant sample damage takes place before the typical imaging time.

\begin{acknowledgments}
The authors ackowledge fruitful discussions with Prof. Marta Fajardo. This work was partially supported by FCT (Portugal) through Grants No. POCTI/FO/FAT/50190/2003 and POCI/FIS/55095/2004.
\end{acknowledgments}

\end{document}